\documentclass[final,1p,times,sort&compress]{elsarticle} 

\usepackage{graphicx}
\usepackage{amssymb} 
\usepackage{amsthm} 
\usepackage{lineno}

\journal{Nuclear Physics A} 

\begin{document}

\begin{frontmatter} 

\title{Quasi-particle and matrix models of the semi Quark Gluon Plasma}

\author[phys,rbrc]{Robert D. Pisarski \fnref{pre}}
\author[rbrc]{Koji Kashiwa}
\author[phys]{Vladimir Skokov}
\address[phys]{Department of Physics, Brookhaven National Laboratory, Upton, NY, 11973 USA}
\address[rbrc]{RIKEN/BNL, Brookhaven National Laboratory, Upton, NY, 11973 USA}
\fntext[pre]{Talk presented by R. D. P.}

\begin{abstract} 
We make a simple observation about two models used to treat
the region near the critical 
temperature of QCD, quasiparticle and matrix models.  While they
appear very different, we show how these two models
might be related.  We also present results for the 
temperature dependence
of the ratio of the shear viscosity to the entropy
in a matrix model, and suggest that quasi-particle models may
behave similarly.
\end{abstract} 

\end{frontmatter} 

\section{Introduction}

The region near the critical temperature in QCD, which can be termed
the semi Quark Gluon Plasma, is
of fundamental interest.  Amongst others, one model used 
to study the semi-QGP is based upon 
quasi-particles
\cite{Peshier:1995ty, Peshier:2005pp, Ozvenchuk:2012fn, Castorina:2011ja};
another, on matrix models
\cite{Pisarski:2000eq, Meisinger:2001cq, Dumitru:2003hp, Hidaka:2008dr, Hidaka:2009ma, Dumitru:2010mj, Dumitru:2012fw, Kashiwa:2012wa, Pisarski:2012bj}.
While these two approaches appear to have nothing in common, 
in this note we suggest how they might be 
related.  We then present a result for
the temperature dependence of the ratio 
of the shear viscosity to the entropy in a matrix model.

A quasi-particle model is based upon assuming that the pressure is given
by that of an ideal gas of massive gluons.
First one computes the pressure for such an ideal gas.
Using lattice results for the pressure, one then determines the gluon
mass.  That is, one trades one function of temperature,
namely the pressure, for another, here the gluon mass. 

\begin{figure}[htbp]
\begin{center}
 \includegraphics[width=0.4\textwidth]{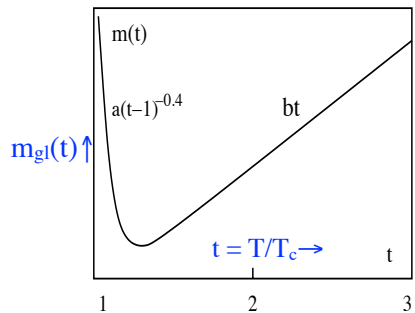}
\end{center}
\caption{The quasi-particle gluon mass from 
Ref. \cite{Castorina:2011ja}, for the pure glue theory with three
colors. }
\label{fig:quasiparticle}
\end{figure}

In Fig. (\ref{fig:quasiparticle}) we show the behavior of the gluon mass
in the pure glue theory with three colors from
Ref. \cite{Castorina:2011ja}, but it is very 
similar to that from Refs. 
\cite{Peshier:1995ty, Peshier:2005pp, Ozvenchuk:2012fn}.
There are two regimes: for temperatures above
$\sim 1.2 \, T_c$, the gluon mass is approximately linear in temperature,
$m_{gl} \sim T$.  This is like the usual Debye mass generated
in perturbation theory.  In fact, its not quite the same,
since in quasi-particle models 
$m_{gl}$ is a mass for transverse excitations.  Still, it is a useful
way of thinking about this gluon mass.

Below $\sim 1.2 \, T_c$, the gluon mass increases strongly, 
as a (fractional) power of $1/(T - T_c)$.  
This sharp increase in the gluon mass near $T_c$ follows inexorably from
the assumptions of the model.  The lattice data indicates that the
pressure of the confined phase is very small 
\cite{Petreczky:2012rq}.  The only
parameter to change in a quasi-particle model is the mass, and so
for an ideal gas of massive particles to give a small pressure, they
must become very heavy, with the pressure strongly
suppressed by Boltzmann factors.  

One thing of note, which is not usually discussed in quasi-particle models,
is clear from Fig. (\ref{fig:quasiparticle}).
Above $1.2 \, T_c$, there is an
an ``ordinary'' quasi-particle regime, where $m_{gl} \sim T$.
In contrast, below $1.2 \, T_c$, there is 
a ``transition'' regime, where $m_{gl}$ increases sharply, driving
the transition to confinement.

Matrix models appear to be very different from quasi-particle models.
One expands about a constant field for the time-like component of
the vector potential, $A_0$.  This is a matrix, albeit extremely simple:
it is diagonal, and constant in space-time.
There are both perturbative terms in $A_0$, and non-perturbative terms,
added by hand 
to drive the transition to confinement
\cite{Meisinger:2001cq,Dumitru:2010mj, Dumitru:2012fw, Pisarski:2012bj, Kashiwa:2012wa}.

In the matrix model, the non-perturbative terms are $\sim T^2$, versus the
usual $T^4$ terms in perturbation theory.  
Instead of fitting to the pressure, one fits to the interaction measure,
$(e - 3p)/T^4$, where $e$ is the energy density.  In the
pure glue theory, this quantity has a narrow peak 
below $\sim 1.2 \, T_c$.  
The parameters of the matrix model are then determined by fitting to
this narrow peak.  Not surprisingly, the details of the matrix model
only matter below $\sim 1.2 \, T_c$.  Thus there is a clear division
between the region about $\sim 1.2 \, T_c$, and that below.  
This is similar to the two regimes in the quasi-particle model,
as seen in Fig. (\ref{fig:quasiparticle}).  After discussing some results
and limits in the matrix model, we return to this similarity below, in 
Eq. (\ref{loop_mass}).

The matrix model was developed in the pure glue theory.  One clean way
of testing the addition of quarks is to use heavy quarks.
In the matrix model, the position of the deconfining critical endpoint
was computed using the matrix model \cite{Kashiwa:2012wa}.
This occurs for a heavy quark mass, about twice 
that for the charm quark, $ \sim
2.4$~GeV for three degenerate flavors.  This result 
agrees well with recent results in a hopping parameter 
expansion \cite{Fromm:2011qi}.  

We now describe some simple computations in a matrix model in the limit
of a large number of colors,
$N_c$.  As for heavy quarks, we add $N_f$ flavors of massless quarks,
simply by adding the quark contribution to the $A_0$ potential
perturbatively.  We assume that $N_c$ is large because then results
for the shear viscosity can be taken directly 
\cite{Hidaka:2008dr, Hidaka:2009ma}.  We give results both for
the pure glue theory, $N_c = \infty$ and $N_f = 0$, 
and for $N_f = N_c = \infty$.

We compute under the
assumption of a uniform eigenvalue density \cite{Dumitru:2012fw}.
This is useful because then the model can be solved trivially.
The exact solution at infinite $N_c$ shows that 
the eigenvalue density is not constant at $N_c = \infty$ 
\cite{Pisarski:2012bj}.  The exact solution exhibits unusual behavior, 
but we do not expect this to affect $\eta/s$ significantly, though.

\begin{figure}[htbp]
\begin{center}
\includegraphics[width=0.4\textwidth]{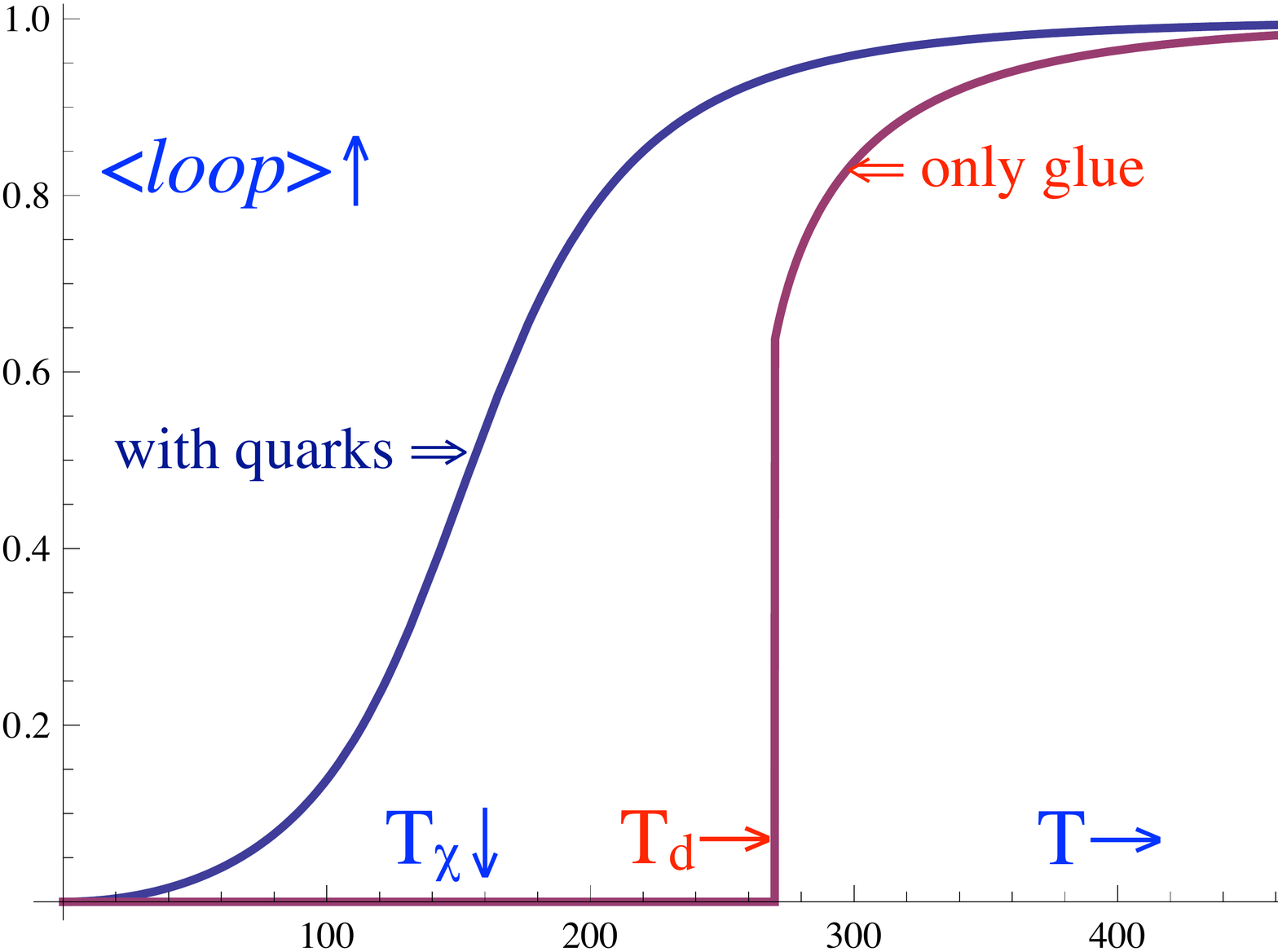}
\hspace{0.1\textwidth}
\includegraphics[width=0.4\textwidth]{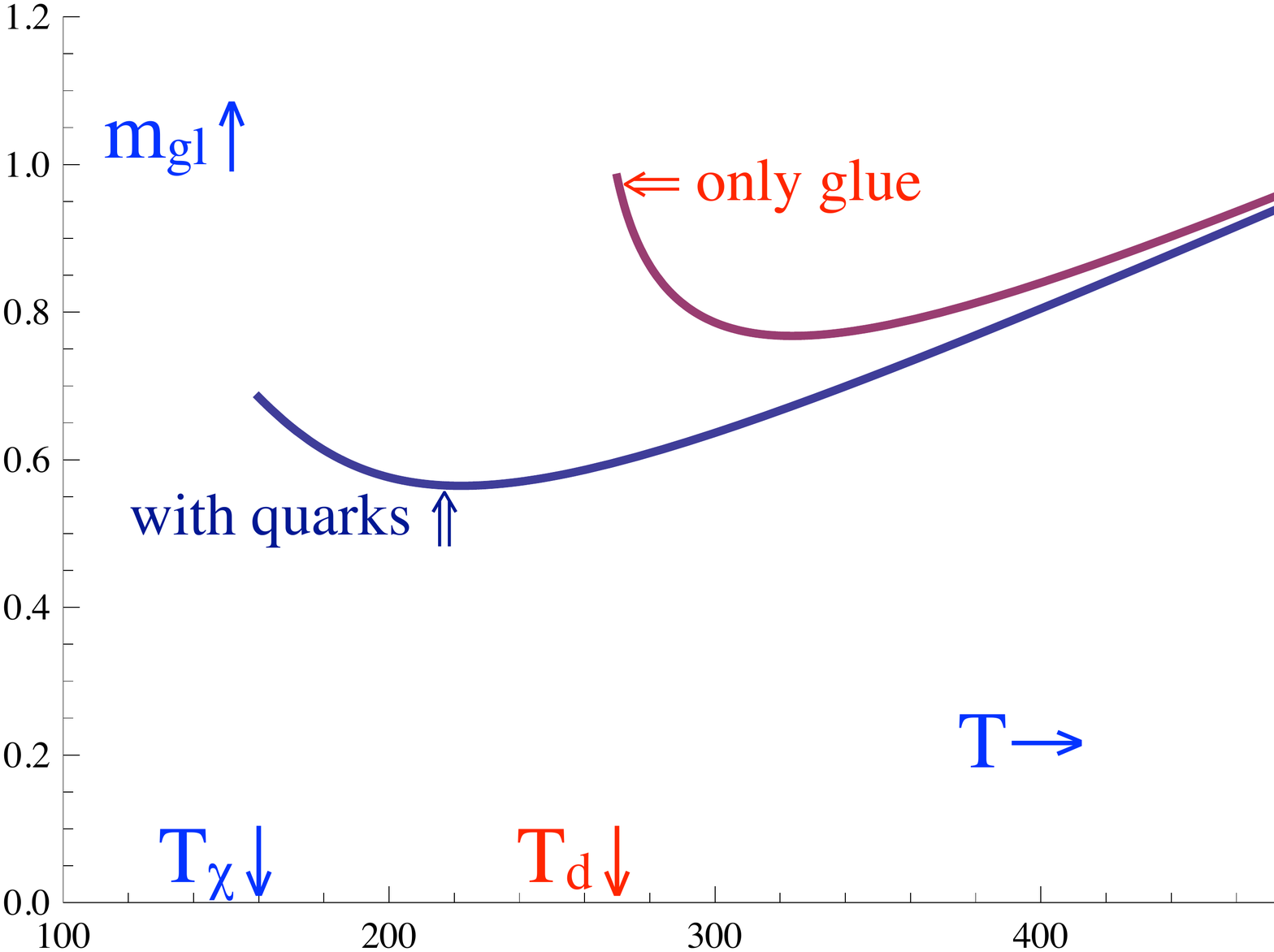}
\end{center}
\caption{Results from matrix models: 
expectation values of Polyakov loops (left) and 
effective gluon masses (right), from Eq. (\ref{loop_mass}). Each
graph includes results for the pure glue theory, $N_f = 0$ and
$N_c = \infty$, and with dynamical quarks, $N_f = N_c = \infty$.
In the right panel, in the pure
glue theory $m_{gl}$ is infinite below $T_d$; with quarks, we choose only to
show it above $T_\chi$, but it continues to increase as the
temperature decreases.}
\label{fig:loops}
\end{figure}

In the left panel of Fig. (\ref{fig:loops}) we show the Polyakov loop,
$\langle loop \rangle = 1/N_c \, {\rm tr} \, {\bf L}$, computed 
for the pure glue theory, and with dynamical quarks.  
(Here ${\bf L}$ is the thermal Wilson line.)  The transition
temperature for the pure glue theory is taken to be as for three colors,
$T_d \sim 270$~MeV.  
With massless, dynamical quarks, we 
{\it assume} there is only a chiral transition,
at $T_\chi$.  We take $T_\chi \sim 160$~MeV from three
colors.  As shown in the left panel,
in the pure gauge theory the loop is zero below $T_d$, while with dynamical
quarks, its expectation value is nonzero at any nonzero temperature.
This figure is very similar to results obtained on
the lattice for three colors: see, {\it e.g.}, Fig. 2 of
Ref. \cite{Petreczky:2012rq}.

We stress that in the matrix model, in principle there is {\it no} strong
relation between $T_\chi$ and $T_d$.  For a small number of flavors,
$N_f \ll N_c$, then $T_\chi \geq T_d$, but when $N_f \sim N_c$, the two
are not correlated, and indeed, $T_\chi$ can be significantly less than
$T_d$.  Physically, this is because as for three colors, many light quarks
simply wash out the deconfining transition.

We now return to the possible similarity between quasi-particle and matrix
models.  As noted, in both models there is a difference between 
temperatures above $\sim 1.2 \, T_c$, and those below.  Let us try to make
this more precise.
The Polyakov loop is the propagator for an infinitely heavy quark.
To represent how it changes, we {\it define} a quasi-particle mass
from the expectation value of the loop,
\begin{equation}
\frac{1}{N_c} \, {\rm tr} \, {\bf L}
= {\rm e}^{ - \, m_{gl}/(\kappa T) } \;\; , \;\; \kappa = 1.7 \; .
\label{loop_mass}
\end{equation}
We stress that this definition of the gluon mass is {\it ad hoc}.  However,
it is not completely
unreasonable.  In the confined phase of a pure gauge theory, the loop
vanishes, and so the gluon mass $m_{gl}$ must be infinite.  The is like
the large (but finite) gluon mass 
in quasi-particle models.  Let is ignore these minor differences, and
concentrate on how the gluon mass
behaves as one approaches the critical temperature from above.

This leads to our basic point.  If we simply
take Eq. (\ref{loop_mass}) as given, and use it to compute the
gluon mass from the left panel of Fig. (\ref{fig:loops}), 
then the gluon masses which results, in the left panel of
Fig. (\ref{fig:loops}), looks rather 
like those of the quasi-particle model, in Fig. (\ref{fig:quasiparticle})!
Admittedly, this agreement depends sensitively upon the value chosen
for the constant $\kappa$ in Eq. (\ref{loop_mass}).  Nevertheless, given
the definition of the gluon mass, the agreement isn't that surprising.
As the loop decreases, the gluon mass must increase.  

The agreement is only qualitative.  Since in the quasi-particle model
the only thing to suppress the pressure is the gluon mass, its increase
below $1.2 \, T_c$, Fig. 1, is much sharper than in the matrix model.
It is amusing that in the matrix model, the gluon mass
appears to be linear in temperature above $1.2 \, T_c$.  This reflects
the narrow transition region in the matrix model, so the loop is
nearly constant.


\begin{figure}[htbp]
\begin{center}
\includegraphics[width=0.4\textwidth]{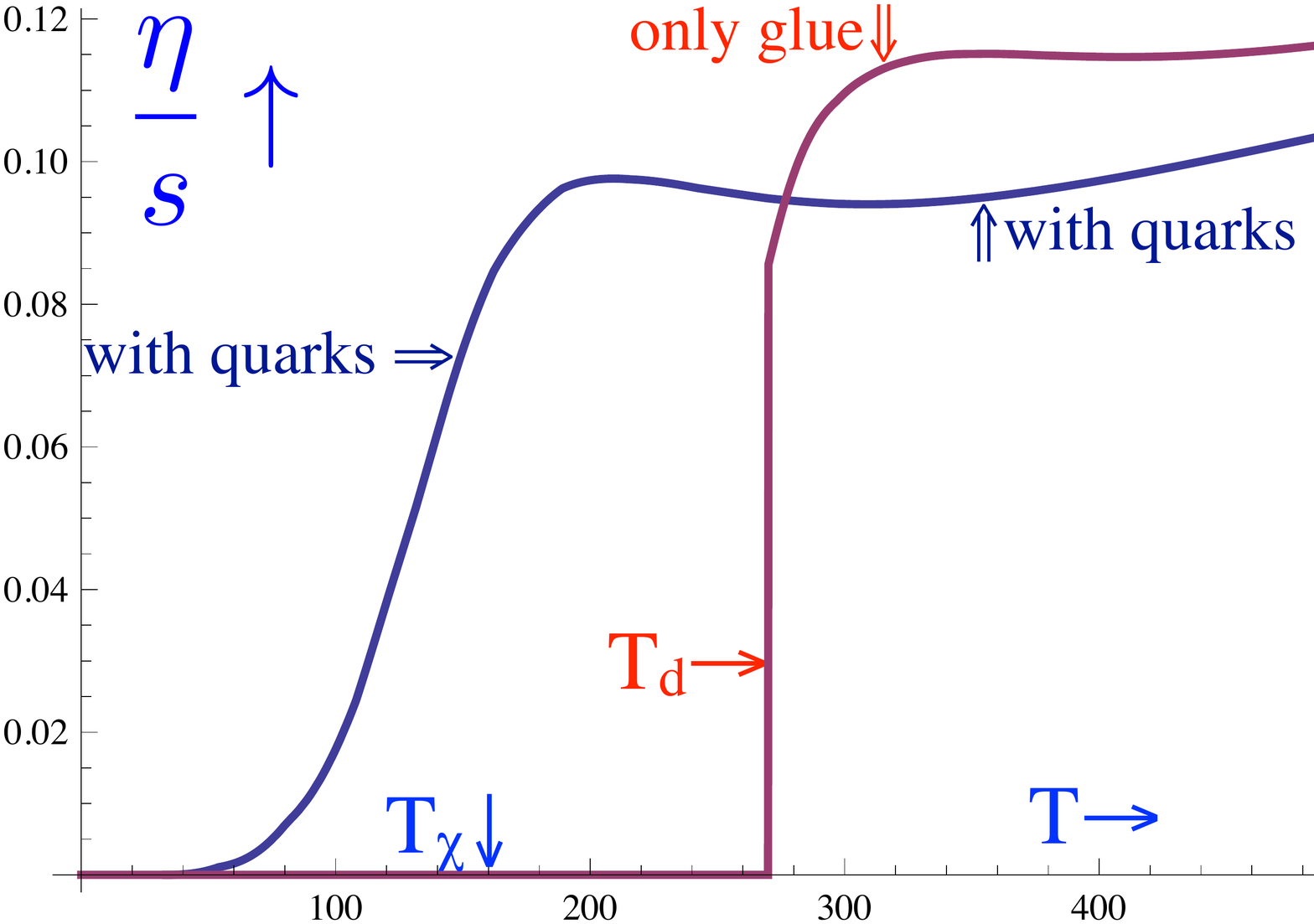}
\end{center}
\caption{The ratio of the shear viscosity to the entropy from matrix models.
Results both with only gluons, with $N_c = \infty$,
and with dynamical quarks, with $N_f = N_c = \infty$, are shown.}
\label{fig:eta_entropy}
\end{figure}

To illustrate what can be done with the matrix model, in 
Fig. (\ref{fig:eta_entropy}) we show the behavior of the ratio of the
shear viscosity to the entropy in the matrix model, following
Ref. \cite{Hidaka:2009ma}.  This computation was done in the limit
of infinite $N_c$, which is why we limited ourselves to this case.

In this computation, various constants have been chosen so that 
the ratio $= 1/(4 \pi)$ at the transition temperature.  This is not
a natural choice, but made out of convenience.  
The point of the exercise, which is meant only as illustrative, is to
show that while $\eta/s$ is certainly temperature dependent,
this can be rather moderate.  By the arguments above,
we suggest that $\eta/s$ may behave similarly in quasi-particle models.

\bibliographystyle{model1-num-names}
\bibliography{QM2012}

\end{document}